\documentclass[aps,nofootinbib,amsmath,amsfonts,preprint]{revtex4}
\usepackage{hyperref}
\usepackage{graphicx,xcolor}
\usepackage{longtable}
\usepackage{subfigure}
\usepackage{xfrac}
\usepackage{comment}
\usepackage{tikz}

\usetikzlibrary{arrows.meta, chains, positioning, shapes.geometric}

\newcommand{\eqr}[1]{Eq.~\eqref{#1}}

\newcommand{\vR}{{\mathbf{r}}}
\newcommand{\vRt}{\vR,\,t}
\newcommand{\dd}{{\mathrm{d}}}

\newcommand{\materialine}[1]{\dd #1/\dd t}

\newcommand{\velocity}{{\mathrm{v}}}

\newcommand{\average}[1]{\left<#1\right>}
\newcommand{\Hamiltonian}{{\mathcal{H}}}

\begin{document}

\title{Non-equilibrium distribution function in ultra-fast processes}

\author{K.~S.~Glavatskiy}
\affiliation{The University of Newcastle, \\University Dr, Callaghan NSW 2308, Australia\\kirill.glavatskiy@newcastle.edu.au}
\date\today

\begin{abstract}
A simple expression for the non-equilibrium distribution function in ultra-fast transient processes is proposed. Postulating its dependence on temporal derivatives of the equilibrium integrals of motion, non-equilibrium analogues of the thermodynamic relationships are derived and the conditions that maximize the non-equilibrium entropy are identified. A rigorous threshold between ``slow" and ``fast" processes is suggested, identifying the range of applicability of classical quasi-equilibrium description. The proposed theory is validated by deriving the known law of inertial heat conduction, which accounts for finite speed of thermal propagation. Finally, a new expression for the non-equilibrium work is derived, revealing two kinds of pressure that emerge in fast non-equilibrium.
\end{abstract}

\maketitle

Statistical description of non-equilibrium systems is typically concerned with response to external perturbations \cite{Balescu} that produce spatial inhomogeneities and corresponding irreversible fluxes, which bring the system back to equilibrium. Systematic descriptions of these processes exist within local \cite{deGrootMazur}, non-local \cite{Jou}, and non-linear \cite{Evans2002} frameworks. A common aspect across these approaches is the assumption of \textit{quasi-equilibrium}, which means that the phase space density distribution function (or simply the distribution function) has the same structure as in equilibrium, i.e. depends on the equilibrium integrals of motion, such as the energy or the number of particles. This is the case irrespective of the strengths of the external perturbation, including arbitrary far-from-equilibrium ones \cite{Jarzynski1997a, Crooks1999}. In other words, in these processes the system has sufficient time to equilibrate (either locally or globally), which essentially implies that these processes are \textit{slow}. 


Here, we develop a new theory that goes beyond the quasi-equilibrium regime, aiming to address \textit{fast} and \textit{ultra-fast} processes. This is manifested in postulating that the non-equilibrium distribution function depends not only on the equilibrium integrals of motion, but also on their temporal derivatives. The latter are non-zero in non-equilibrium and it is reasonable to expect that they contribute to the system's temporal evolution. 
We will refer to this regime as \textit{transient equilibrium} or simply ``non-equilibrium".

Consider a system with the distribution function $\rho \equiv \rho(\Gamma(t))$, where $\Gamma$ is the set of all microscopic coordinates and $\Gamma(t)$ is the system's temporal microscopic trajectory in its phase space. For any function $x = x(\Gamma)$, we will use the argument $\Gamma$ to indicate that it is a microscopic function that depends on the microscopic coordinates $\Gamma$, and we will use angle brackets $\average{x} \equiv \average{x(\Gamma)}_{\Gamma}$ for its phase space average. Furthermore, we will use $\materialine{x(\Gamma)}$ or $\dot{x}(\Gamma)$ to denote the temporal derivative of $x$ that acts on its microscopic coordinates, as well as any external parameters that $x$ may depend on. Finally, we will use the Hamiltonian symbol $\Hamiltonian(\Gamma)$ to denote all equilibrium integrals of motions, i.e. the energy $E(\Gamma)$ \textit{and} the number of particles $N(\Gamma)$. In particular, with this notation, we can write for the system in global equilibrium $\dot{\rho}_{eq} = 0$ and $S_{eq} = - \average{\ln\rho_{eq}}$, where $S$ is the entropy of the system.

We write the following identity for the equilibrium distribution function
\begin{equation}\label{eq/01}
\ln\rho_{eq}(\Gamma(t)) = \Big[1 + \tau\,\frac{\dd }{\dd t}\Big]\,\ln\rho_{eq}[\Hamiltonian(\Gamma(t))]
\end{equation}
where the operator $\tau\,(\materialine{})$ represents a shift along the microscopic trajectory by an arbitrary time $\tau$. \eqr{eq/01} is trivially satisfied since the equilibrium distribution function remains unchanged along its microscopic trajectory, according to Liouville's theorem \cite{ll5}. In particular, because the equilibrium distribution function depends only on the integrals of motion, the energy $E(\Gamma)$, and the number of particles $N(\Gamma)$ (or their equivalents in each quantum state, if considering a quantum system), which are constant in equilibrium, the second term in \eqr{eq/01} is zero.

Consider now the system that undergoes an arbitrary fast process such that $\dot{E} \neq 0$ or $\dot{N} \neq 0$. As in a quasi-equilibrium process, its statistical description is governed by two contributions: thermal, which accounts for exchange of energy or particles with the thermostat, and mechanical, which accounts for external work. We do not consider spatial relaxation processes here, which result in internal fluxes.  

The thermal contribution is governed by the distribution function. We postulate that the non-equilibrium distribution function $\rho(\Gamma)$ has the following structure:
\begin{equation}\label{eq/03}
\ln\rho(\Gamma(t))= \Big[1 + \tau\,\frac{\dd }{\dd t}\Big]\,\ln\rho_{eq}[\Hamiltonian(\Gamma(t))]
\end{equation}
In contrast to \eqr{eq/01}, the shift time $\tau$ here is no longer arbitrary. It depends on the specific details of microscopic interactions in the system (e.g., the form of the Hamiltonian), and within the proposed theory must remain as a phenomenological parameter. 

\begin{figure}[h]
\begin{tikzpicture}[
node distance = 5mm and 7mm,
start chain = going right,
alg/.style = {align=center, font=\linespread{0.8}\selectfont}
]
\begin{scope}[every node/.append style={on chain, join=by -Stealth}]
\node (n4) [alg] {\textit{full}\\ \textit{equilibrium} \\\\  \\state \\\\ $\rho_{eq}[\Hamiltonian(\Gamma)]$ \\\\ $\tau=0$};
\node (n3) [alg] {\textit{quasi} \\ \textit{equilibrium} \\\\ slow \\process \\\\ $\rho_{eq}[\Hamiltonian(\Gamma(t))]$ \\\\ $\tau=0$};
\node (n2) [alg] {\textit{transient}\\ \textit{equilibrium} \\\\ fast \\process \\\\ $\rho[\Hamiltonian(\Gamma(t)),\dot{\Hamiltonian}(\Gamma(t))]$ \\\\ $\tau \neq 0$};
\node (n1) [alg] {\textit{no}\\ \textit{equilibrium} \\\\  dynamical \\system  \\\\ $\Gamma(t)$ \\\\ };
\end{scope}
\end{tikzpicture}
\caption{Hierarchy of (non-)equilibriums: as the speed of the process increases, more variables are required for its description. }
\label{fig/01}
\end{figure}
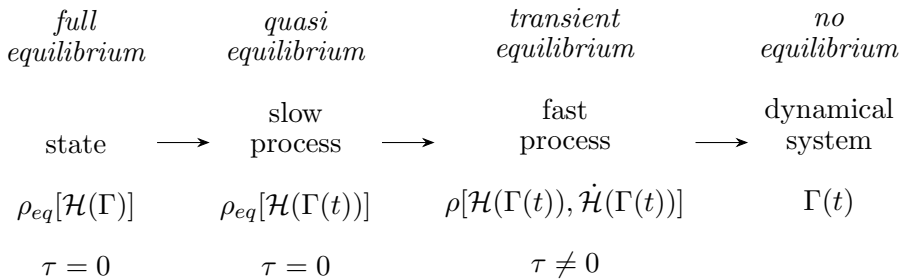

It is important to emphasize that the mere use of a distribution function in non-equilibrium implies possibility of a ``reduced" description, so that instead of the complete set of all microscopic coordinates, the system can be characterized by a small number of some collective quantities. This puts a certain ``upper bound" on the types of non-equilibrium processes that can be addressed with the help of a distribution function, creating a certain hierarchy of processes with respect to their speed and appropriate description, as illustrated in Figure~\ref{fig/01}. With this perspective, $\tau$ can be interpreted as a measure of the system's equilibration time. For $t \gg \tau$ the system is in quasi-equilibrium, and its distribution function ${\rho(\Gamma(t)) = \rho_{eq}[\Hamiltonian(\Gamma(t))]}$. In contrast, for $t \ll \tau$ the system is in transient equilibrium, and its distribution function ${\rho(\Gamma(t)) = \rho[\Hamiltonian(\Gamma(t)),\,\dot{\Hamiltonian}(\Gamma(t))]}$. Consequently, the distinction between ``slow" and ``fast" processes can be made with the reference to $\tau$: the process is fast if its characteristic time is less than $\tau$, and slow otherwise. In particular, for many systems $\tau = 0$, which means that they equilibrate essentially instantaneously, and therefore the processes they participate in should be characterized as ``slow", so that their evolution can be described by regular approaches of statistical mechanics. In contrast, for systems with non-zero $\tau$, the transient effects become important, which the present work aims to address. We will see below that $\tau$ is inversely proportional to the speed of propagation of thermal perturbations; within equilibrium statistical mechanics, such speed is infinite, which is equivalent to $\tau=0$. 
Finally, when there is ``no equilibrium", a ``reduced" description of the system in terms of the distribution function is limited. The system in this case can either be described by a position-dependent distribution function $\rho_{eq}[\Hamiltonian(\Gamma(\vRt))]$ (corresponding to local, non-local, non-linear response theories mentioned above) or may not allow any reduced description at all (corresponding to dynamical systems); this case is not discussed here.

The mechanical contribution is governed by external forces that change ``generalized coordinates" $\lambda$. These coordinates contribute to the system's Hamiltonian as parameters but may change due to the corresponding ``generalized force". Without loss in generality, we will refer here to a single pair of the generalized ``force-coordinate" as ``pressure-volume", using $\lambda$ and $V$ interchangeably; the analysis can be directly generalized to any other pair. Indicating the explicit dependence of the microscopic Hamiltonian on $\lambda$, we write $\dot{E}(\Gamma,\,\lambda) = (\partial E/\partial \lambda)\,\dot{\lambda}$. In quasi-equilibrium, the derivative in front of $\dot{\lambda}$ determines the pressure $p_{eq} \equiv - \average{\partial E/\partial \lambda}_{eq}$, which is equal to the external pressure (or other forms of external ``force"); we will refer to $\pi(\Gamma,\,\lambda) \equiv - \partial E/\partial \lambda$ as the microscopic pressure. Consequently, $\dot{E}$ would depend on both $\lambda$ and $\dot{\lambda}$, so the variation of $\dot{E}$ due to a change of the external parameter can be written as 
\begin{equation}\label{eq/05}
\dd \dot{E}(\Gamma,\,\lambda,\,\dot{\lambda}) = - \pi(\Gamma,\,\lambda)\,\dd \dot{\lambda} - \frac{\partial\pi(\Gamma,\,\lambda)}{\partial\lambda}\dot{\lambda}\,\dd\lambda
\end{equation}

Consider now the explicit form of the Gibbs distribution function for the grand canonical ensemble. In quasi-equilibrium, it can be written as
\begin{equation}\label{eq/06}
\ln\rho_{eq}(\Gamma)= \frac{1}{T_{eq}}\,\Big[\Omega_{eq}-E(\Gamma)+\mu_{eq}\,N(\Gamma)\Big]
\end{equation}
where $T_{eq}$ and $\mu_{eq}$ are the quasi-equilibrium temperature and chemical potential respectively, while $\Omega_{eq}$ is the quasi-equilibrium grand potential, which also serves the role of the normalization factor for the distribution function. Substituting \eqr{eq/06} in \eqr{eq/03}, we obtain the following expression for the non-equilibrium distribution function
\begin{equation}\label{eq/07}
\ln\rho(\Gamma)= \frac{1}{T}\,\Big[\Omega-E(\Gamma)+\mu\,N(\Gamma) - \tau_{_E}\,\dot{E}(\Gamma) + \mu\,\tau_{_N}\,\dot{N}(\Gamma)\Big]
\end{equation}
where $T$ and $\mu$ are the non-equilibrium temperature and chemical potential respectively, while $\Omega$ is the normalizing factor which can be interpreted as the non-equilibrium grand potential. Since the relaxation dynamics of the energy and the particles may differ, we have explicitly introduced two different shift times $\tau_{_E}$ and $\tau_{_N}$. Compared with the quasi-equilibrium distribution function, the non-equilibrium distribution function has additional terms proportional to the shift times; we will refer to such terms as \textit{transient}. 

We define the non-equilibrium entropy in the standard way
\begin{equation}\label{eq/08}
S \equiv - \average{\ln\rho(\Gamma)}_\Gamma 
\end{equation}
so the grand potential
\begin{equation}\label{eq/09}
\Omega = - TS + \average{E} - \mu\average{N} + \tau_{_E}\dot{\average{E}} - \mu\tau_{_N}\dot{\average{N}}
\end{equation}
has, compared to the quasi-equilibrium expression, additional transient terms too. Taking the differential of the normalization condition, $\average{\rho(\Gamma)}_\Gamma = 1$, and using \eqr{eq/05} with $\lambda$ as the volume $V$, we obtain for the differential of the grand potential
\begin{equation}\label{eq/10}
\begin{array}{rl}
\dd\Omega =& -S\,\dd T - \average{N}\,\dd\mu - P\,\dd V 
\\
&- p\,\tau_{_E}\,\dd \dot{V} - \dot{\average{N}}\,\dd(\mu\tau_{_N}) + \dot{\average{E}}\,\dd\tau_{_E}
\end{array}
\end{equation}
where we have introduced
\begin{equation}\label{eq/26}
\begin{array}{rl}
p \equiv& \average{\pi}
\\
P \equiv& \average{\pi} + \tau_{_E}\,\average{\frac{\partial\pi}{\partial V}}\dot{V}
\end{array}
\end{equation}
Taking the differential of \eqr{eq/09} and subtracting \eqr{eq/10} from it, we obtain the non-equilibrium Gibbs relation
\begin{equation}\label{eq/11}
\begin{array}{rl}
T\,\dd S = &\dd \average{E} + P\,\dd V - \mu\,\dd\average{N} 
\\
&+ \tau_{_E}\,\dd\dot{\average{E}} + p\,\tau_{_E}\dd \dot{V} - \mu\,\tau_{_N}\dd\dot{\average{N}}
\end{array}
\end{equation}

\eqr{eq/11}  provides the basis for deriving all necessary non-equilibrium thermodynamic relations for transient non-equilibrium processes. In particular, the suite of non-equilibrium thermodynamic potentials can be introduced, as in quasi-equilibrium, via Legendre transformations, which must, however, include the transient terms. Indeed, repeating the above procedure for the canonical ensemble, the non-equilibrium free energy is introduced as the normalizing factor in the canonical distribution function, resulting in ${F = - TS + \average{E} + \tau_{_E}\dot{\average{E}}}$, which includes the transient term. Thus, the Gibbs free energy ${G \equiv F - \Omega = \mu\,[\average{N} +\tau_{_N}\,\dot{\average{N}}]}$ and the enthalpy ${H \equiv G + TS = TS + \mu\,[\average{N} +\tau_{_N}\,\dot{\average{N}}]}$  include the transient terms as well. The non-equilibrium thermodynamic potentials, including the entropy, must be additive, i.e., scale linearly with the system's size. This automatically translates to the rates of change of the corresponding variables: e.g., for a system consisting of two subsystems 1 and 2, ${\dot{(E_1+E_2)} = \dot{E}_1+ \dot{E}_2}$ and ${\dot{(N_1+N_2)} = \dot{N}_1+ \dot{N}_2}$. Note that such additivity does not apply to the speed of the process: the non-equilibrium thermodynamic potentials should not scale e.g. up by a factor of 2 if the rate of the volume change is increased by a factor of 2. This means that the non-equilibrium potentials are first degree homogeneous functions of their extensive variables, just like in quasi-equilibrium. It follows from \eqr{eq/10} that the extensive variables, which $\Omega$ depends on, are $V$ and $\dot{V}$ only, so 
\begin{equation}\label{eq/12}
\begin{array}{rclcl}
\Omega &=& \Omega\;(T,\,V,\,\mu,\;\;\tau_{_E},\,\dot{V},\,\tau_{_N}\mu) &=& - P\,V - p\,\tau_{_E}\,\dot{V}
\\
G &=& G\;(T,\,P,\,N,\;\tau_{_E},\,\tau_{_E}p,\,\dot{N}) &=& \mu\,\average{N} +\mu\,\tau_{_N}\,\dot{\average{N}}
\\
F &=& F\;(T,\,V,\,N,\;\tau_{_E},\,\dot{V},\;\;\;\dot{N}) &=& G + \Omega
\\
E &=& E\;(S,\,V,\,N,\;\dot{E},\,\;\dot{V},\;\;\;\dot{N}) &=& G + \Omega + TS 
\\
H &=& H\;(S,\,P,\,N,\,\tau_{_E},\,\tau_{_E}p,\,\dot{N}) &=& G + TS 
\end{array}
\end{equation}

It is straightforward to show that the distribution function $\eqref{eq/07}$ maximizes the entropy $\eqref{eq/08}$, provided both $\average{E}$, $\average{N}$ and $\dot{\average{E}}$, $\dot{\average{N}}$ are fixed. Such situation, when both $E(\Gamma)$, $N(\Gamma)$ and $\dot{E}(\Gamma)$, $\dot{N}(\Gamma)$ fluctuate around some given values, corresponds to fast oscillations of $E(\Gamma)$ or $N(\Gamma)$. This means that a fast oscillating dynamical system is actually in full \textit{thermodynamic equilibrium}, but this equilibrium is described by additional transient variables $\dot{\average{E}},\,\dot{\average{N}},\,\dot{V}$. 
If, however, the transient dynamics is such that $E(\Gamma)$ or $N(\Gamma)$ change fast but arbitrarily (e.g., monotonically), then the distribution function $\eqref{eq/07}$ does not maximize the entropy $\eqref{eq/08}$, and the system undergoes an arbitrary fast non-equilibrium process. The present analysis is applicable to either case.

We next discuss two specific examples of fast transient processes: inertial heat conduction and non-equilibrium work.

\paragraph{Inertial heat conduction.}

Consider two macroscopic rigid bodies of the same size and material at different temperatures, $T_1 > T_2$, that are brought in contact with each other. As long as the temperatures are different, there exists an irreversible heat flux $J$ between the bodies, which carries energy from the warmer body to the cooler one. In the classical description (that implies a slow process), the heat flux follows Fourier's law of heat conduction, ${J_{F} = -K\,(T_1-T_2)}$, where $K$ is the thermal conductance. The Fourier's expression for the heat flux ensures that the total entropy production of the system, $\dot{S} = \dot{S}_1 + \dot{S}_2$, is always positive, reaching zero when the bodies' temperatures become equal. For slow processes, the total entropy production is $\dot{S} = -J_{F}\,(T_1-T_2)/(T_1\,T_2)$, which follows from the simple energy balance 
\begin{equation}\label{eq/21}
\begin{array}{rl}
\dot{\average{E}}_{1} = -J
\\
\dot{\average{E}}_{2} = +J
\end{array}
\end{equation}
and the quasi-equilibrium Gibbs relation (i.e. \eqr{eq/11} with $\tau_{_E} = \tau_{_N} =0$) with constant volume and number of particles.

For fast processes, the Gibbs relation \eqref{eq/11} is different from the quasi-equilibrium one, which results in a different expression for the energy flux $J$. In particular, for the system considered
\begin{equation}\label{eq/22}
T_{1,2}\,\dot{S}_{1,2} = \dot{\average{E}}_{1,2} + \tau_{_E}\,\ddot{\average{E}}_{1,2}
\end{equation}
where the double dot indicates the second derivative with respect to time and subscripts $1,2$ indicate that \eqr{eq/22} applies to both bodies. The energy balance is still described by \eqr{eq/21}; taking its time derivative and substituting both to \eqr{eq/22}, the total entropy production becomes
\begin{equation}\label{eq/23}
\dot{S} = - (J + \tau_{_E}\,\dot{J})\,(T_1-T_2)\,/(T_1\,T_2)
\end{equation}
which is different from the quasi-equilibrium one in the flux factor: $J + \tau_{_E}\,\dot{J}$ instead of $J_{F}$. To ensure that the entropy production is positive, i.e., the second law of thermodynamics is not violated, one must require that 
\begin{equation}\label{eq/24}
J + \tau_{_E}\,\dot{J} = - K\,(T_1-T_2)
\end{equation}
which is the famous Maxwell-Cattaneo-Vernotte (MCV) equation for fast heat transfer, where $\tau_{_E}$ is referred to as the relaxation time or the so-called thermodynamic inertia \cite{Glavatskiy2025}. Note, that in the form \eqref{eq/24}, this equation is applicable to heat transfer not only between spatially neighboring bodies, but also between components that occupy the same spatial position (e.g., in the so-called ``two-temperature model"). This case is discussed in detail in \cite{Glavatskiy2025}.

The relaxation time $\tau_{_E}$ in \eqr{eq/24} is exactly the same quantity as the shift $\tau_{_E}$ in \eqr{eq/03}. Expressing the energy change as $\dot{\average{E}} = c_{\velocity}V\dot{T}$, where $c_{\velocity}$ is the volumetric heat capacity, and using \eqr{eq/21} and \eqr{eq/24} to write the differential equation for the temperature, it can be shown that for times $t \ll \tau_{_E}$ this differential equation has the form of the wave equation with the propagation speed $u = \sqrt{\kappa/(c_{\velocity}\tau_{_E})}$, where $\kappa$ is the thermal conductivity \cite{Jou}. Thus, the shift time is 
\begin{equation}\label{eq/25}
\tau_{_E} = \kappa/(c_\velocity u^2)
\end{equation}
If the thermal propagation speed is infinite, then the shift time is zero; this corresponds to the classical description of non-equilibrium thermodynamics or the quasi-equilibrium regime. In contrast, any finite thermal propagation speed implies non-zero shift time; this corresponds to the conditions beyond classical non-equilibrium and thus implies dependence of the distribution function on the temporal derivatives of the equilibrium integrals of motion. This explains the physical meaning of the shift time.

\paragraph{Non-equilibrium work.}

The microscopic force with which the system acts on its boundary at position $\mathbf{r}$ is calculated from mechanics as $\mathbf{F} = - \partial E(\Gamma,\mathbf{r})/\partial \mathbf{r}$, so the work performed by the system when the generalized coordinate $\lambda$ (which depends on $\mathbf{r}$) changes by $\dd\lambda$ is $\dd A = \average{\mathbf{F}}|_{_S}\cdot\dd\mathbf{r} = - \average{\partial E/\partial \lambda}|_{_S}\,\dd\lambda$, where the phase space averaging is performed under adiabatic conditions. This expression is valid both in equilibrium and in non-equilibrium. When the generalized coordinate is the volume, the work performed by the environment on the system (which is opposite in sign to the work performed by the system) is 
\begin{equation}\label{eq/27}
\dd A = - p\,\dd V
\end{equation}
where $p$ defined by the first equation of \eqr{eq/26}. \eqr{eq/27} simply reflects the mechanical (rather than thermodynamic) definition of work, with $p$ being the external force applied to the system per unit of its surface area. 

At the same time, the $P\,\dd V$ term, which contributes to the non-equilibrium Gibbs relation \eqref{eq/10}, is different from the work \eqref{eq/27}. Indeed, $P$ is given by the second equation of \eqr{eq/26} and is different from $p$. This suggests that one can identify two different quantities with the meaning of pressure, $p$ and $P$. It may be useful to interpret $p$ as the ``external" pressure (as it contributes to the work performed by or against the environment), while $P$ as the ``internal" pressure (as it contributes to the thermodynamic potentials of the system) or, alternatively, as ``mechanical" and ``thermodynamic" pressure respectively. In a {quasi-equilibrium process}, $\tau_{_E} = 0$, so $P = p = p_{eq}$, and the work performed on the system by the environment coincides with the corresponding term in the Gibbs relation. In contrast, in a non-equilibrium process, while the system itself is in equilibrium (the transient one), it is not in equilibrium with the environment. Indeed, the difference between these pressures is 
\begin{equation}\label{eq/28}
\frac{P-p}{\rho_m c^2} = - \tau_{_E}\,\frac{\dot{V}}{V} 
\end{equation}
where $\rho_m$ is the material density, $c$ is the speed of sound in the media (which is different from the speed of thermal propagation), and we have used the expression for the adiabatic compressibility $\beta_{_S} \equiv - V^{-1}\,(\partial p/\partial V)|_{_S} = 1/(\rho_m c^2)$. It is known that mechanical equilibration happens much faster than thermal one. \eqr{eq/28} suggests that the former takes a time of the order of $\tau_{_E}$. If $\tau_{_E} = 0$ (which is the case when thermal equilibration is described by Fourier's law), then mechanical equilibration is indeed instantaneous. In contrast, if $\tau_{_E} \neq 0$, the system requires this additional ``catch-up" time, to reach mechanical equilibrium with the environment.

Finally, it is instructive to provide the expression for work in an adiabatic process for a system with fixed number of particles, i.e. when $\dd S = 0$, $\dd\average{N}=0$, $\dd\dot{\average{N}}=0$. In quasi-equilibrium, this results in a simple form of energy conservation: $\dd A_{eq} = \dd\average{E}$, i.e. the external work is converted to the internal energy in full. In non-equilibrium (i.e. when $\tau_{_E} \neq 0$), however, it follows from \eqr{eq/11} that
\begin{equation}\label{eq/28}
\dd A|_{_S}  = \dd \average{E} + (P-p)\,\dd V + \tau_{_E}\,(\dd\dot{\average{E}} + p\,\dd\dot{V})
\end{equation}
Thus, in addition to the internal energy change, the external work is converted to several transient terms.

To summarize, we have proposed a new theory of statistical description of fast and ultra-fast processes. The theory does not rely on any specific microscopic details of the system and hence should be applicable to a broad range of real systems. The analysis in this paper offers a clear distinction between slow and fast processes and thus identifies the range of applicability of traditional quasi-equilibrium frameworks. This, in particular, explains why quasi-equilibrium approaches have been successful in the description of many non-equilibrium processes: it is because the system's shift time is negligible compared with the characteristic time of the process, in which the system participates. For ultra-fast processes this is not necessarily the case, hence the predictions of this theory should be directly measurable in such processes. In particular, the proposed theory provides a rigorous explanation of inertial heat transport, which is observed in ultra-fast processes \cite{Jou}, and explains why mechanical equilibration is much faster than thermal one. Among other predictions, the theory offers non-equilibrium analogues of thermodynamic relationships, including the Gibbs relation and the expression for non-equilibrium work. This paper opens a possibility for rigorous description of a broad range of fast and ultra-fast processes.

\bibliographystyle{unsrt}
\bibliography{irreversible}

\end{document}